\documentstyle[aps,twocolumn]{revtex}
\begin{document}
\title{The Symmetries of Fermion Fluids at Low Dimensions}

\author{Philip W. Anderson and F. Duncan M. Haldane}
\address{Joseph Henry Laboratories of Physics\\
Princeton University, Princeton, NJ 08544}

\leftline{\bf Abstract}

We point out that the quasiparticle spectrum of the Landau Fermi liquid
theory has
an extra $Z_2$ symmetry, local in momentum space, which is not generic to
the
Hamiltonian with interactions.  Thus the Fermi liquid is in this sense a
(quantum)
zero-temperature critical point.


For many decades it has been accepted that the ``natural" low-temperature
state of a
fermion fluid in the absence of a magnetic field is the Fermi Liquid
described by
Landau.$^{(1)}$
Landau's fundamental assumption was that the states of an interacting
electron gas (we
shall hereafter call the fermions electrons) could be put into a one-to-one
correspondence
via adiabatic continuation with those of the free electron gas, an
assumption which
appears at first sight very plausible and is the basis of several
derivations of the
Fermi liquid theory.

Parenthetically, we note that it is accepted that in fact this is never
actually the case
at very low temperature, where a logarithmic divergence appears in one or
another
``Cooper channel" $k\sigma, -k-\sigma \to k'\sigma', -k' -\sigma'$ and
deforms the state
into a BCS state for $D \geq 2$. But repulsive interactions are weakened by
the Cooper
channel divergence, while the most important attractive interactions, those
due to
phonons, are effective only at low energies, so that often the Fermi liquid
has a wide
range of validity.

We remark here that the Fermi liquid is, from a symmetry point of view,
surprising, in
that the Fermi liquid has an extra symmetry which is generically broken by
the
interactions. Only if the interaction term renormalizes to zero at low
temperature is the
Fermi liquid in principle the correct solution.

The symmetry of the kinetic energy terms for a free Fermi gas embodies the
fact that there
are separately conserved currents of both signs of the spin at every point
in space--or,
if we wish, separately conserved currents at each momentum.  Landau's famous
argument,
following Weisskopf and others, is that scattering terms leave this
conservation intact
at the Fermi momentum due to exclusion  principle prohibitions on final
states. There are
thus conserved currents of ``up" and ``down" spins where the quantization
direction
defining ``up" is arbitrary at each momentum. There are thus two degenerate
conserved complex
fermions at each $k_F$, which we may describe in terms of four real
(Majorana) fermions
if we like.

Thus the fundamental symmetry is the group of real rotations among four
objects, i.e.
$O(4)$. $O(4)$ is isomorphous with the direct product of two SU(2)'s and the
discrete
group $Z_2$. $Z_2$ occurs because the rotations of $O(4)$ may be divided
into two classes,
proper and improper rotations, depending on whether the determinant of the
rotation is
$+1$ or $-1$. The simplest improper rotation just reverses the sign of one
real fermion,
which is equivalent to an electron--hole transformation for one spin.  Thus
$O(4)$ may be
divided by $Z_2$, and the resulting group of proper rotations is
$$
O(4) \div Z_2 \ =\ SU(2) \times  SU(2)
$$
The two separate $SU(2)$ symmetries thus revealed are those of conserved
charge current
and conserved spin current.  The idea of charge as an $SU(2)$ is a little
unfamiliar
but is behind the Anderson-Nambu Pauli operators $\tau_1, \tau_2, \tau_3$
used in the BCS
theory to describe the charge degrees of freedom of singlet pairs.$^{(2)}$
The kinetic energy,
for particles not at $E_F$, multiplies $\tau_3$ and reduces the charge
symmetry to
$U(1)$.

The short-range repulsive interaction $Un_\uparrow n_\downarrow$ maximally
violates the
extra $Z_2$ symmetry of the Landau theory: it obviously changes sign under
improper
rotations, i.e. under hole-particle transformations for one spin only.  This
is an
operation which interchanges charge and spin. But since charge and spin are
conserved at
every collision, the separate $SU(2)$ symmetries should be retained in the
interacting
fluid.

It has been assumed that $U$ doesn't break the $Z_2$ symmetry of the Fermi
liquid, since,
as we said above, at the Fermi level the quasiparticle currents are
perfectly conserved
because of the exclusion principle, and apparently the effects of $U$ have
become
irrelevant. But this is not the whole story: $U$ modifies the density of
states and the
response coefficients, and in fact we know that the Landau interaction
coefficients are
not the same for charge and spin: the compressibility and spin
susceptibility are not
related as they are for free particles because they contain Landau mean
field corrections,
and zero sound has a different velocity from spinwaves.

The assumption is, however, that at the level of single elementary
excitations we still
have the $O(4)$ or $U(2)$ symmetry with the extra $Z_2$. In $1D$ it is known
that this
assumption fails: the Landau Fermi liquid gives way to the Luttinger
liquid,$^{(3)}$
which has
only the separated charge and spin symmetries and has two separate
velocities of elementary
excitations. Analysis of the reason for this shows us that there is an
interaction
term which fails to renormalize to irrelevance in this case, namely the
forward scattering
phase shift for opposite--spin particles of zero relative momentum.  This
term effectively
causes a chiral anomaly: the two currents of up and down spins are no longer
separately
conserved, rather there is a mixing of up-spin hole current with down-spin
particles and
vice versa.

In the conventional theory of the Fermi liquid in 3D this term in fact does
renormalize
to irrelevance because of the large phase space for recoil momenta: it is
possible to
renormalize the simple Born approximation by a simple partical wave analysis
to make
the scattering phase shift vanish proportionally to the relative momentum
Q.$^{(4)}$
$$ \eta_0 \ =\ Q a$$
in 3 dimensions,  where $a$ is the effective scattering length used,
especially, in the low
density calculations of Huang and coworkers.$^{(5)}$

2D is the critical dimensionality for this renormalization.  This case was
examined by
Bloom$^{(6)}$ and he showed that in the low-density limit $a$ diverges as
$1/Q|\ell n Q|$,
which gives
a barely convergent theory.  But as we have shown before, at any finite
density the effective
length theory does not renormalize the phase shift to zero and $U$ can
remain marginally
relevant at the level of quasiparticle energies.

Thus in one and two dimensions the Fermi liquid can be thought of as a
quantum critical
point with an
extra $Z_2$ symmetry, which is exact only at the singular point $U = 0$ (or
$n = 0$). In
general, no phase transition is required into the charge-spin separated
state.  In higher
dimensionality, the Fermi liquid acquires the $Z_2$ symmetry continuously as
$T \to 0$,
which again is a kind of quantum critical point.

It is suggestive to think of the localized spin$^{(7)}$ as a model for the
above process.
At $T >>
U$, the electron gas is effectively free and there is no local spin.  As we
go into the
region $U > T >> T_K$, this renormalizes into the Kondo model, which has a
free electron
gas interacting with a local spin (i.e. the interacting  part of the gas has
now separate
charge and spin dynamics.)

Finally, as $T < T_K$ this renormalizes continuously away and we return to a
quasiparticle
resonance with no separate spin dynamics. The analogy is not perfect but it
is suggestive.
As Zlatic has shown,$^{(8)}$ the low-temperature fixed point is analytically
continuable from
the high-temperature non-interacting state, yet the intermediate $\tau$
states have no
resemblance to either.

In conclusion, we have shown that the generic symmetry of the electron fluid
involves
separately conserved charge and spin currents.  The quasiparticle concept
implies a larger
symmetry which in $> 2$ dimensions manifests itself in the Fermi liquid,
which is a $T =
0$ quantum critical point.  For 1D and perhaps 2 the $T = 0$ state need not
have this
extra symmetry.

It is a pleasure, especially for the first author, to submit this paper to a
volume in
honor of Quin Luttinger, who was an old and valued friend and a superb
physicist.


\leftline{REFERENCES:}

\leftline{(1) L.D. Landau, Soviet Physics JETP 3, 920 (1956).}

\leftline{(2) P.W. Anderson, Phys. Rev. 112, 1900 (1958),}
\leftline{\quad Y. Nambu, Phys. Rev. 117, 648 (1960).}

\leftline{(3) F.D.M. Haldane, J. Phys.  C14, 2585 (1981).}
\leftline{\quad E. Lieb and F.Y. Wu, Phys. Rev. Lett. 20, 1447 (1967).}

\leftline{(4) K. Huang and C.N.Yang, Phys. Rev. 105, 767 (1958).}

\leftline{(5) V.M. Galitskii, Soviet Physics JETP 7, 104 (1958).}

\leftline{(6) P. Bloom, Phys. Rev. B12, 125 (1975).}

\leftline{(7) F.D.M. Haldane, Phys. Rev. Lett. 40, 416 (1978),}
\leftline{\quad P.W. Anderson, Phys. Rev. 124, 1 (1959).}

\leftline{(8) V. Zlatic and B. Horvatic, Phys. Rev. B28, 6904 (1983).}

\end{document}